\documentclass[compsoc,conference,a4paper,10pt,times]{IEEEtran}
\IEEEoverridecommandlockouts
\usepackage{amsmath,amssymb,amsfonts}
\usepackage{graphicx}
\usepackage{xcolor}
\usepackage{enumitem}
\usepackage{amssymb}
\usepackage{footnote}
\usepackage{pifont}
\usepackage{makecell}
\usepackage{booktabs}
\usepackage{mathtools}
\usepackage{algorithm}
\usepackage{algpseudocode}
\usepackage{subcaption}
\usepackage{multirow}
\usepackage{url}
\usepackage{gensymb}
\usepackage{balance}
\usepackage{stfloats}
\usepackage{cite}
\usepackage{amsmath,amssymb,amsfonts}
\usepackage{graphicx}
\usepackage{bmpsize}
\usepackage{xcolor}
\usepackage{lipsum}
\usepackage{textcomp}
\usepackage[colorlinks=true,urlcolor=black]{hyperref}
\def\BibTeX{{\rm B\kern-.05em{\sc i\kern-.025em b}\kern-.08em
    T\kern-.1667em\lower.7ex\hbox{E}\kern-.125emX}}

\newcommand{\myparagraph}[1]{\vspace{4pt} \noindent \textbf{#1.}}
\makesavenoteenv{tabular}
\makesavenoteenv{table}

\DeclarePairedDelimiter\abs{\lvert}{\rvert}
\DeclarePairedDelimiter\norm{\lVert}{\rVert}
\algnewcommand\algorithmicinput{\textbf{INPUT:}}
\algnewcommand\INPUT{\item[\algorithmicinput]}

\newcommand{\knowledge}{limited knowledge }

\begin{document}

\title{Biometric Backdoors: A Poisoning Attack Against Unsupervised Template Updating\\
\thanks{This paper was accepted at the 5\textsuperscript{th} European Symposium on Security and Privacy 2020.}  
}

\author{\IEEEauthorblockN{Giulio Lovisotto,
		Simon Eberz and Ivan Martinovic}
	\IEEEauthorblockA{University of Oxford\\
		Email: firstname.lastname@cs.ox.ac.uk}}

\maketitle

\begin{abstract}
In this work, we investigate the concept of biometric backdoors: a template poisoning attack on biometric systems that allows adversaries to stealthily and effortlessly impersonate users in the long-term by exploiting the template update procedure.
We show that such attacks can be carried out even by attackers with physical limitations (no digital access to the sensor) and zero knowledge of training data (they know neither decision boundaries nor user template).
Based on the adversaries' own templates, they craft several intermediate samples that incrementally bridge the distance between their own template and the legitimate user's.
As these adversarial samples are added to the template, the attacker is eventually accepted alongside the legitimate user.
To avoid detection, we design the attack to minimize the number of rejected samples.

We design our method to cope with weak assumptions for the attacker and we evaluate the effectiveness of this approach on state-of-the-art face recognition pipelines based on deep neural networks.
We find that in white-box scenarios, adversaries can successfully carry out the attack in over 70\% of cases with less than ten injection attempts.
Even in black-box scenarios, we find that exploiting the transferability of adversarial samples from surrogate models can lead to successful attacks in around 15\% of cases.
Finally, we design a poisoning detection technique that leverages the consistent directionality of template updates in feature space to discriminate between legitimate and malicious updates.
We evaluate such a countermeasure with a set of intra-user variability factors which may present the same directionality characteristics, obtaining equal error rates for the detection between 7-14\% and leading to over 99\% of attacks being detected after only two sample injections.
\end{abstract}


\section{Introduction}
In recent years, biometric authentication has become one of the preferred ways to mitigate burdens associated with passwords (e.g., re-use, inconvenient creation policies and bad memorability).
With a long history of research, face and fingerprint recognition are the most popular modalities and authentication systems based on them are commonly delivered with consumer products.
While early research focused on the performance of these modalities under a zero-effort threat model, current trends in biometric systems are also prioritizing high protection of biometric \textit{templates}, i.e., the user's stored biometric information.
Templates in fact represent sensitive user data and their leak might compromise the secrecy of the biometric trait in a permanent way.
Different measures can be used to protect user templates: standard file encryption, template protection algorithms (e.g., cancelable biometrics~\cite{Patel2015}, key generation~\cite{Nandakumar2008}), secure elements~\cite{faceidsecurityguide} and distributed template storage~\cite{fidouaf}.
Nowadays, most commercial biometric systems use any combinations of these protection measures.
As an example, both Apple's TouchID and FaceID store templates in the protected secure enclave and the data never leaves the user's device~\cite{faceidsecurityguide}.

Due to these protection mechanisms, attackers can not modify the template directly, since that would require them to compromise the device.
However, an adversary can exploit the \textit{template update} procedure to cause the system to adapt the template, thereby indirectly modifying the data within.
In fact, during template update, the system replaces or adds recently seen samples to the user template.
Template update allows systems to cope with the inherent variability of biometric traits, such as physiological changes or sensor measurement noise, by collecting additional user samples.

Biometric systems use either supervised or unsupervised updating strategies. 
Supervised updating means that additional guarantees in the user identity are required before the update (e.g., Windows Hello~\cite{WinHello} requires the users to re-input their PIN or password before template update).
From a usability perspective, unsupervised updating represents a more desirable choice, as it is a seamless approach that does not require additional user interaction.
In both scenarios, the confidence in the user identity can be strengthened with the introduction of a \textit{self-update} threshold: a new sample for the update is discarded if it does not match a confidence threshold (e.g., needs to be sufficiently similar to the current user template).
For example, FaceID automatically augments the user template with a new sample after a successful authentication if the sample quality (i.e., low  noise) is sufficient~\cite{faceidsecurityguide}.

By exploiting this template update procedure, an attacker can  inject adversarial samples into the user template to carry out a  \textit{template poisoning} attack.
In this attack, the adversary corrupts the user template allowing himself to impersonate the user with his own biometric trait.
The poisoned template is hard to detect and creates an inconspicuous and stealthy backdoor for the adversary in the long-term.
Once placed, the backdoor allows the adversary to access the system without requiring them to modify their appearance.
However, the adversary needs to overcome four key challenges: (i) he has limited control over the injected samples, (ii) all injected samples must clear the update threshold, (iii) he has limited knowledge of the authentication system and the legitimate user's template, (iv) he must avoid degrading the legitimate user's experience to avoid generating suspicion.

In our analysis, we focus on face recognition as it is one of the most well-known and widely used biometric modalities, including in unsupervised environments.
Faces show inherent variability caused by changes in lighting environment, sensor position and user behavior.
We present a method to carry out template poisoning attacks on biometric systems based on deep machine learning pipelines.
We evaluate our attack on state-of-the-art systems for face recognition: we show that leveraging population data is sufficient to make the attack successful even for adversaries with limited knowledge and capabilities.
While we consider a white-box scenario where the feature extraction network and the update policy is known, we assume that the adversary does not know the user template and does not have digital access to the sensor.
We also show that even relaxing the known network assumption, i.e., in a black-box case, the attack can still be carried out with some success.
Afterwards, we propose a new countermeasure for the detection of poisoning attacks based on the angular similarity of samples.
Compared to previous work~\cite{biggio2015adversarial, Kloft2010}, we also evaluate the detection taking into account the trade-offs with legitimate template updates, showing that it can effectively stop poisoning attacks.

\myparagraph{Contributions}
The contributions of this paper are as follows:
\begin{itemize}[leftmargin=0.6cm]
	\item We propose a method to plant \textit{biometric backdoors} by poisoning the legitimate user's template in \knowledge (known network, update policy and pupulation data, unknown user template) and limited injection capabilities scenarios.
	\item We evaluate the attack on state-of-the-art recognition pipelines, including white- and black-box models. We show that the error rates of the system hardly change when such a backdoor is present, making the attack inconspicuous.
	\item We introduce a poisoning detection method that thwarts poisoning attacks without affecting legitimate template updates, and we investigate these trade-offs on a large face dataset.
\end{itemize}

\section{Related Work}
\label{sec:relatedwork}

In this section, we first give a brief background on biometric recognition, we then discuss attacks on biometric systems, and then given an overview of the existing work in poisoning attacks.

\subsection{Biometric Recognition}
Both the research community and industry have shown significant interest in biometric recognition in recent years.
Often times, recognition of biometric traits is proposed as an approach to mitigate the shortcomings of passwords (such as bad memorability, password reuse and easy to guess passwords).
Previous research has shown that most biometrics present inherent intra-user variability, both for physiological and behavioral traits.
In the following we will give a brief overview of face recognition systems, and how intra-user variability affects the system.

\myparagraph{Face Recognition}
Modern face recognition systems are based on deep convolutional neural networks (DNN), where the input image undergoes transformation over several convolutional layers (e.g., FaceNet~\cite{Schroff}, VGG-Face~\cite{Parkhi15}, ResNet~\cite{cao2018vggface2}).
In recent years, state-of-the-art models have shown to outperform humans in recognition accuracy.
Most face recognition DNN can also be used as feature extractors, allowing them to work with faces unseen in the training data.
In this case, another simpler classifier is added onto the recognition pipeline that uses the output of the neural network as its inputs.

\myparagraph{Template Update}
The physiological trait (face) itself does not show significant intra-user variations, but when using image-based recognition, a series of factors may influence its appearance to the sensor.
Well known variation factors include age, pose (viewing angle), lighting environment, facial hair and facial accessories.
Some of these variations can be accounted for at enrolment, for example asking the user to rotate their head as some samples are collected or shining additional light onto the users face to account for the effect of external lighting conditions. 
However, systems also increasingly rely on template updating.
As an example, both Apple's FaceID~\cite{faceidsecurityguide} and TouchID~\cite{touchidsupportpage} perform template update procedures.
FaceID performs template updates either after a new sample is accepted, or when a rejected sample is followed promptly by a correctly entered backup passcode.

\subsection{Adversarial ML Attacks}
Adversarial machine learning has become extremely relevant due to the wide-spread use of deep neural networks, which are prone to being fooled by purposefully crafted adversarial samples.
Adversarial ML attacks have been classified into two main categories: \textit{evasion} (\textit{inference-time}) attacks and \textit{poisoning} (\textit{training-time}) attacks~\cite{papernotsok2018}.
In evasion attacks, the adversary attempts to craft a sample that is classified as belonging to a desired output class.
In poisoning attacks, the adversary injects adversarial samples into the training data in order to maximize an objective function (typically to cause misclassifications at test time).

\myparagraph{Realizability of Adversarial Attacks}
Adversarial samples have been initially investigated in the digital domain, where the adversary is able to create pixel-perfect modifications to the inputs.
Such perturbations rarely survive in the physical world, as a series of factors affects the sensor measurements. As a result, these modifications are not evident to the underlying recognition system. 
For a camera, factors such as viewing angle, resolution, distance have all been shown to affect the measurements enough to severely harm the effectiveness of adversarial samples.
However, recent studies have shown that adversarial samples can also be constructed to survive in the physical world.
As an example, Sharif et al~\cite{Sharif2016} have shown how a carefully crafted frame of glasses can be worn by an individual to dodge face detection or even impersonate arbitrary users.
Their work is further extended in~\cite{Sharif2017}, where the authors use generative adversarial networks to improve the success rate and the transferability across models of the attack.
Other known real-world attacks include adversarial patches~\cite{brown2017adversarial,karmon2018lavan} or posters and graffiti~\cite{Eykholt2017, song2018physical}, where maliciously printed images can be applied on top of objects so that the sensors inputs contain the added image, in order to deceive the classifier (either object or traffic sign classification).

\myparagraph{Poisoning Attacks}
Differently from evasion, in poisoning attacks the goal is to modify the model to misbehave on specific inputs that are known by the adversary without compromising its performance on regular inputs~\cite{Gu2017,Liu2018}.
In biometric systems, poisoning attacks may be categorized into two different categories: (i) poisoning the DNN, (ii) poisoning the user template.
When the DNN is the target of the poisoning, there is no assumption on controlling the DNN inputs, but rather the adversary has an increased control over the training phase, i.e., can directly edit the network weights~\cite{Liu2018} or change the training data~\cite{Gu2017}.
However, as DNNs are increasingly used as black-box feature extractors in a classification pipeline rather than directly as classifiers, analysis of poisoning of the template database needs to be included in the security analysis.
As mentioned in the introduction, template update represents an opportunity for malicious samples to reach the template database.
A preliminary analysis of template poisoning attacks on biometric systems was proposed by Biggio et al.~\cite{Biggio2012}.
The authors present a poisoning attack where an adversary injects a set of samples that gradually shifts the user template towards the biometric trait of the adversary.
The same authors extended their work reducing the knowledge assumptions of the adversary, showing that the attack is feasible even considering a black-box scenario and limited knowledge of the victim template~\cite{Biggio2013}.

In the following we motivate how this paper builds on the assumptions of these works to account for the challenges that an adversary would face in practice in carrying out a poisoning attack.


\begin{table}[tpb]\small
	
	\centering
	\begin{tabular}{c|@{\hskip0pt}c@{\hskip0pt}|@{\hskip0pt}c@{\hskip0pt}|@{\hskip0pt}c@{\hskip0pt}|@{\hskip0pt}c@{\hskip0pt}|@{\hskip0pt}c@{\hskip0pt}|@{\hskip0pt}c@{\hskip0pt}|@{\hskip0pt}c@{\hskip0pt}}
		\thead{Work} & \rotatebox[origin=l]{90}{\thead{\footnotesize{template update}}} & \rotatebox[origin=l]{90}{\thead{\footnotesize{limited injection capabilities}}} &  \rotatebox[origin=l]{90}{\thead{\footnotesize{unknown template}}}  & \rotatebox[origin=l]{90}{\thead{\footnotesize{state-of-the-art recognition}}} & \rotatebox[origin=l]{90}{\thead{\footnotesize{unknown network (black-box)}}} &
		\rotatebox[origin=l]{90}{\thead{\footnotesize{stealthiness}}} &
		\rotatebox[origin=l]{90}{\thead{\footnotesize{uncompromised training phase}}} \\
		\toprule
		
		Kloft et al.~\cite{Kloft2010} & \checkmark & \ding{55} & --- & --- & \ding{55} & --- & \checkmark \\
		\midrule
		
		Biggio et al.~\cite{Biggio2013} & \checkmark & \ding{55}  &  \ding{55} & \ding{55}  &  \checkmark & \ding{55} & \checkmark \\	
		\midrule
		Biggio et al. ~\cite{Biggio2012} &\checkmark & \ding{55} & \ding{55} & \ding{55} &  \ding{55} & \ding{55} & \checkmark \\
		\midrule
		Garofalo et al.~\cite{garofalo2018fishy} & \ding{55} & \ding{55} & \ding{55} & \checkmark & \ding{55} & \ding{55} & \ding{55} \\
		\midrule
		Liu et al.~\cite{Liu2018} & \ding{55} & ---  & \checkmark & \checkmark & \ding{55} & \checkmark & \ding{55}* \\
		\midrule
		This paper & \checkmark & \checkmark& \checkmark & \checkmark& \checkmark & \checkmark  & \checkmark\\
		\bottomrule
	\end{tabular}
	
	\small{\vspace{.25cm}*network weights are changed after training}
	\caption{ Comparison with previous work. A dash (---) indicates that the property does not apply directly to the work. }
	\label{tab:comparison}
\end{table}

\subsection{Differences from Previous Work}
We use the previous work in physically-robust adversarial samples as one of the building blocks of our work, which allows us to control and carry out the poisoning attack as a whole.
We report a comparison to previous work in Table~\ref{tab:comparison}, where we focus on the following properties.

\myparagraph{Template Update}
We consider that template updates are limited by self-update thresholds.
However, we consider penalties for consecutive rejected authentication attempts.
This is a reasonable security policy as sequences of failed authentication attempts might correspond to attackers trying to impersonate the user, and falling back to a different authentication method can thwart such attempts (e.g., FaceID allows five failed attempts before switching to PIN input).
In comparison, previous works~\cite{Biggio2012, Biggio2013, Kloft2010} did not consider or investigate the frequency of rejected attempts.

\myparagraph{Limited Injection Capabilities} 
We consider only cases that account for the physical realizability of the adversarial sample creation and injection. Adversarial examples have been shown to be realizable in the real-world~\cite{Eykholt2017, Sharif2016, Kurakin2016}, but their flexibility and effectiveness is greatly reduced in that case. Assuming perfect injection capabilities, where the adversary can freely inject samples and \textit{digitally} manipulate exactly each pixel
in the input is an unrealistic assumption in our scenario.
In fact, being able to feed arbitrarily manipulated samples to the system, which bypass liveness detection (such as the images in~\cite{garofalo2018fishy, Biggio2012, Biggio2013}) would mean that the adversary could simply be able to authenticate inconspicuously at any time just by having a printed image of the user.
This defeats the purpose of carrying out a more complex poisoning attack.

\myparagraph{Unknown Template}
Previous work focused on full template knowledge~\cite{Biggio2012, garofalo2018fishy}.
Biggio et al.~\cite{Biggio2013} considered a partial knowledge scenario where the known sample is chosen as the closest to user's centroid out of \textit{all} the samples in the testset.
In a biometric system, this assumption would require the adversary to have availability of several samples from the user, which might not be easily obtainable.
Therefore, we focus on an adversary that does \textit{not} have any knowledge of the template stored in the user device, but only has a single image of the user.

\myparagraph{Unknown Network}
While we start our analysis with a white-box scenario, we extend it to black-box ones using the principle of transferability of adversarial samples~\cite{szegedy2013intriguing, goodfellow2014explaining, papernotsok2018}.
A black-box attack was shown to be feasible in~\cite{Biggio2013}, however the system relied on non-state-of-the-art face recognition methods (i.e., EigenFaces~\cite{turk1991eigenfaces}) that are nowadays greatly outperformed by deep neural networks.

\myparagraph{Stealthiness}
We consider a poisoning attack to be \textit{stealthy} when it can be carried out without compromising the normal error rates of the system.
This is a desirable property as changes in false acceptances or false rejections (i.e., legitimate user cannot authenticate, other users can authenticate) would be suspicious: users might stop using the biometric system or re-enroll their template in an attempt to reset the system performance.
This approach separates us from previous work, such as~\cite{Biggio2013,Biggio2012,garofalo2018fishy}, which aims to \textit{replace}, rather than alter the legitimate user's template, blocking them out of the system.
With enrolment being a procedure that the user can carry out independently, such attacks would not guarantee long-term access to the adversary, as  victims would re-enrol into the system as soon as they are inconvenienced by the reduced recognition rates.

\myparagraph{Uncompromised Training Phase}
Unlike previous work, we assume no control over training data~\cite{garofalo2018fishy} or network weights~\cite{Liu2018}.
While these assumptions may lead to more effective attacks, they also require the adversary to be able to affect the network training process.
In our scenario, it is reasonable that a network would be trained beforehand, and would not be accessible once on the device (e.g., Apple FaceID is stored and executed in a secure element~\cite{faceidsupportpage}).

These properties lead our experimental design and evaluation. 
Our goal is to bridge the gap between a purely theoretical poisoning attack and the challenges that an adversary would have to face in practice, allowing us to better understand the feasibility of such attacks in the real-world.

\section{System and Threat Model}\label{sec:systemandthreatmodel}
\begin{figure}[t]
	\centering
	\includegraphics[width=0.48\textwidth]{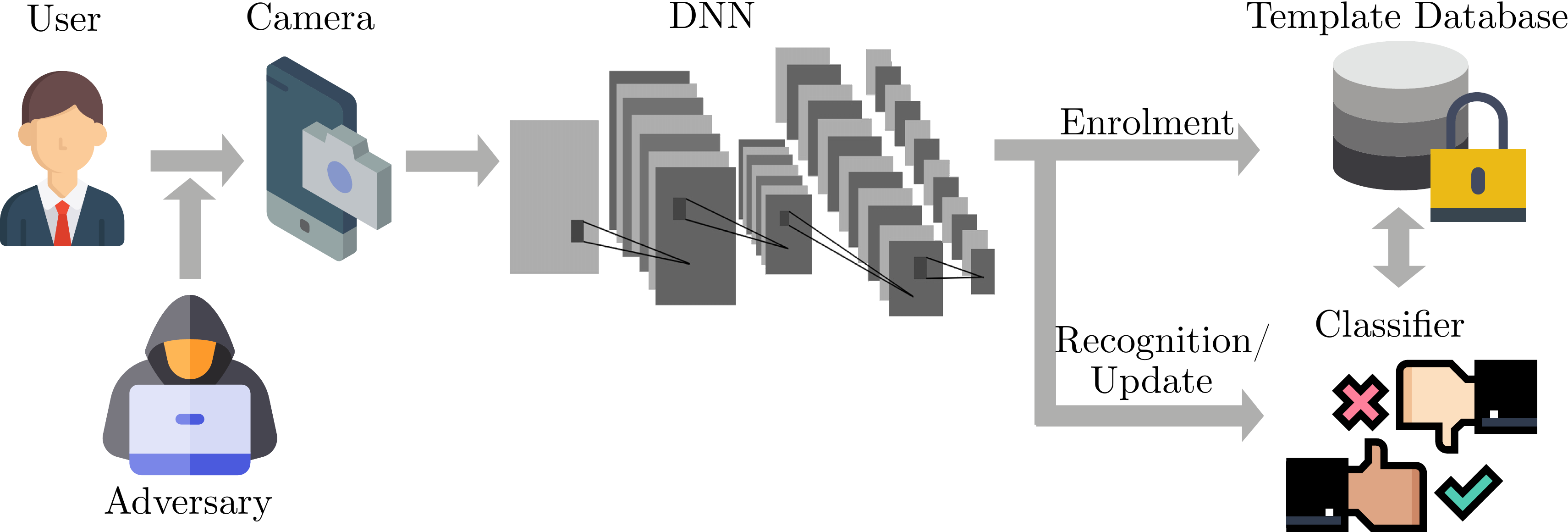}
	\caption{System and threat model. The adversary has physical realizability constraints, meaning that the adversarial samples need to survive through the sensor measurement. Captured images undergo a feature extraction step. 
		The samples' features are then used for enrolment, and a classifier processes them during recognition and update.}
	\label{fig:threat-model}
\end{figure}
In this section we first give an overview of the system model and later talk about the considered adversary.

\subsection{System Model}
\label{sec:systemmodel-overview}
Modern biometric recognition has four phases: development, enrolment, recognition and template updating.

\myparagraph{Development}
During development, a deep neural network is trained on a large labelled dataset of biometric samples.
The objective of the DNN is to learn how to compute features from user samples so that the extracted features lie in a  reduced-dimensionality linearly separable hyperspace.
The extracted features can be either \textit{embeddings}~\cite{Schroff} when triplet loss is used for training, or they can be the features extracted from the layer preceding the last softmax layer (\textit{logits}).
This pre-trained network is then deployed with the biometric system.


\myparagraph{Enrolment}
At enrolment, the users present their biometric traits to the sensor.
One or more measurements are taken, processed and fed into the DNN.
The resulting samples are stored in the template database, and are known as the \textit{user template}.
These samples are kept in a secured space in the user device and are used as a reference whenever an authentication attempt is made.
An illustration of this process is given in Figure~\ref{fig:threat-model}.

\myparagraph{Recognition/Update}
During the recognition phase, whenever a user presents his trait to the sensor, the system measures it, feeds it into the DNN to extract the features and attempts to match the resulting sample and the known biometric template with a threshold-based classifier. 
If the match occurs, the user is successfully authenticated.
In the update procedure the system might decide to add the newly seen sample into the template database if it is ``confident enough'' (based on the similarity to enrolment data) that the sample belongs to the legitimate user.
There are scenarios where the update procedure might use a more conservative threshold than the authentication procedure i.e., of the samples that pass authentication, only those that are very similar to the known template will be added to it.
However, current authentication systems generally set the same threshold for both procedures in order to efficiently keep up with changes in the user template (e.g.,~\cite{personalized-hey-siri, faceidsecurityguide}), we therefore focus on this scenario.

\subsection{Adversary Model}
The adversary's overarching goal is to place a ``biometric backdoor'' which grants them stealthy long-term access to the secured system without requiring further effort once the backdoor is in place. In this section, we define the attacker's objectives, knowledge and capabilities. 

\myparagraph{Objectives}
The attacker's goals are to:
\begin{itemize}[leftmargin=0.5cm]
	\item Cause modifications of the user template that leads to the attacker being accepted.
	\item Maintain the stealthiness of the attack, i.e., minimize the changes to false rejects (FRR) and false accepts (FAR).
	\item Minimize the number of physical accesses to the system required to plant the backdoor.
	\item Minimize the number of samples rejected by the system.
\end{itemize}

\myparagraph{Capabilities}
The adversary has physical access to the sensor and can therefore feed samples to it. 
The adversary does not have digital access to the sensor, as in that case he would be able to perfectly and effortlessly control the inputs, making a poisoning attack unnecessary.
The adversary can therefore alter biometric sensor measurements only in the physical domain, and these will be subject to both sensor and presentation noise.
Unlike previous work, we do \textit{not} assume that the attacker has access to the system during enrolment.
The adversary is also unable to directly change the template or training data (e.g., by removing or replacing user samples).

\myparagraph{Template Knowledge}
In line with related work, we assume that the adversary has at least one picture of the user's face.
Being a physiological feature, face appearance is notoriously difficult to keep secret.
Social media in particular is a plentiful source for videos or photographs~\cite{Xu2016}.
Differently from previous work, where the best matching picture was chosen out of the test data~\cite{Biggio2013}, we do not  set any requirement on the known picture other than it being accepted by the system.
This assumption entails two obstacles for the adversary:  he does not know enrolment data nor the resulting decision boundaries of the system's classifier.

\myparagraph{System Knowledge}
As our goal is to evaluate the effectiveness of poisoning attacks and construct possible defenses, we first assume a strong attacker with white-box access who has access to the DNN used to extract the embeddings after it has been trained.
The adversary additionally knows what algorithms are being used by the recognition pipeline i.e., the type of threshold-based classifier and the template update threshold.
We will show how the adversary can use population data (ideally publicly available, e.g. for face, ~\cite{cao2018vggface2, Huang2007}) to learn certain distribution properties of users' embeddings distances.
We further investigate a black-box attacker using the principle of transferability of adversarial samples~\cite{papernot2016transferability}. 
In this case, the adversary has knowledge of a surrogate model (which replaces the DNN) that he can use to optimize the attack on the black-box network.

\section{Attack Concept}

\label{sec:attackconcept}
Using the assumptions made in the system and threat model, we present an overview of the attack and of its challenges.

\subsection{Overview}

The concept behind the poisoning attack is that the adversary adds adversarial samples to the legitimate user template in order to change the decision boundary of the classifier.
Figure~\ref{fig:poisoning-example} shows a two-dimensional representation of how the attack works.
There are three categories of samples:
\begin{itemize}[leftmargin=0.5cm]
	\item \textit{user (victim) samples}: legitimate user samples;
	\item \textit{attacker samples}: samples coming from the biometric trait of the adversary;
	\item \textit{poisoning samples}: samples algorithmically crafted by the adversary.
\end{itemize}
Figure~\ref{fig:poisoning-example} shows how the user and attacker samples are well separated in the feature space, due to the uniqueness of their biometric traits.
At enrolment, the classifier learns the distribution of the user samples creating a boundary around it, shown by the darker blue area.
The classifier is able to correctly discriminate between attacker and user samples, rejecting the adversary in an impersonation attempt.

Knowing his own template and a user sample as the starting point, the adversary crafts the poisoning samples accordingly.
As the self-update threshold is in place, the adversary must make sure that the crafted samples lie within the current accepted region (shaded blue area), otherwise they would be rejected as anomalous.
By injecting one poisoning sample at a time, the adversary shifts the decision boundary towards his own sample distribution.
With sufficient poisoning samples, the adversary will move the decision boundary enough so that his own samples will fall inside it, and can therefore impersonate the user with his own trait.


\subsection{Challenges}

\begin{figure}[t]
	\centering
	\includegraphics[width=0.425\textwidth]{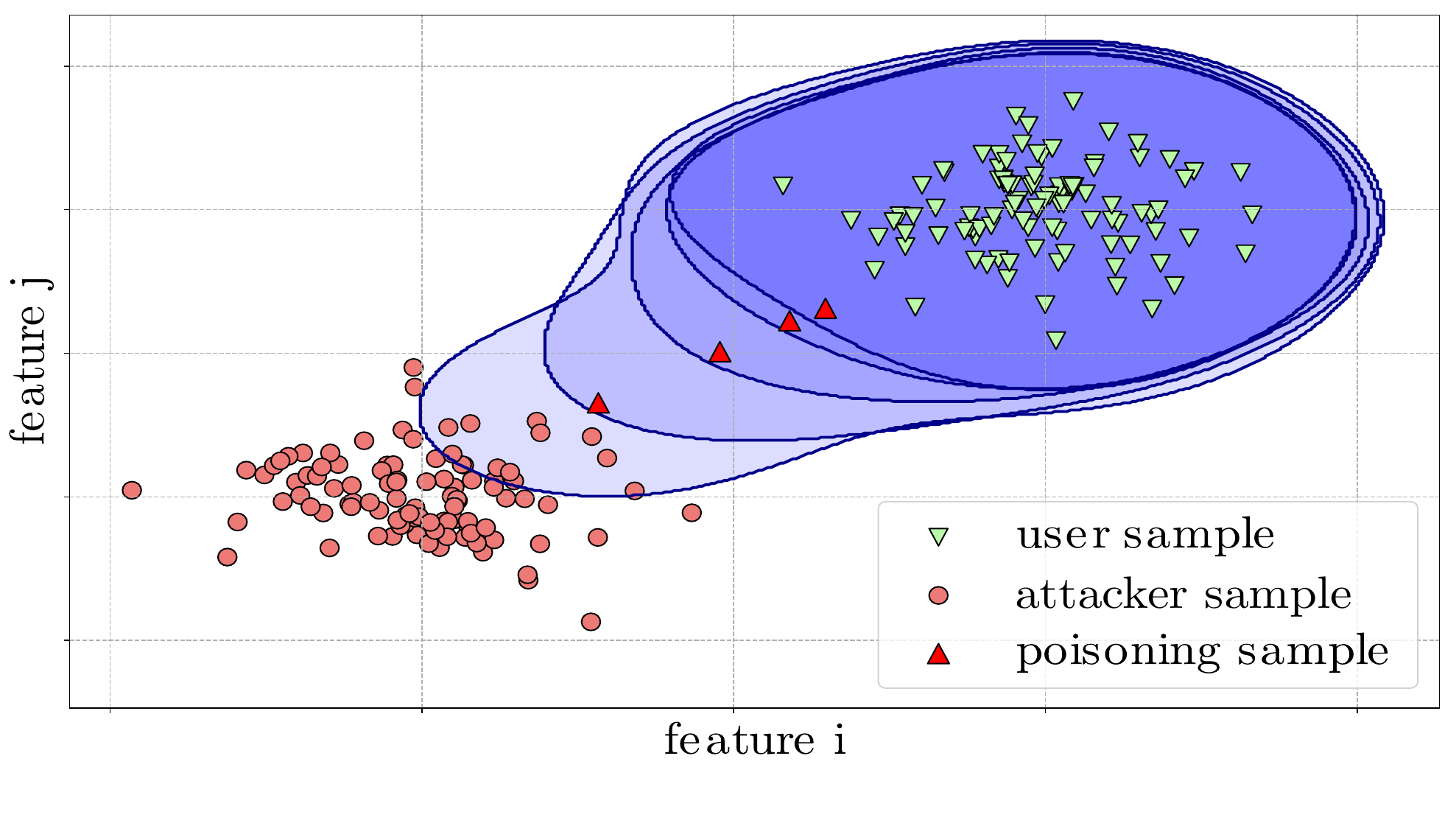}
	\caption{Changes in the classifier decision boundary (shaded areas) with the addition to the template of adversarial crafted samples. With enough poisoning samples, the classifier will recognize the adversary as the legitimate user.}
	
	\label{fig:poisoning-example}
\end{figure}

The challenges in designing the attack are based on the limited injection capabilities (which bring input uncertainty) and on the unknown user template and location of decision boundaries (which cause failures in the injection).

\myparagraph{Input Uncertainty} Figure~\ref{fig:poisoning-example} shows a simplified 2-dimensional example of the user distribution.
This distribution is actually in an \textit{n}-dimensional feature space that consists of hundreds of dimensions.
If the adversary had exact control over the location of samples in this space, he could simply craft and inject the poisoning samples that lie on the \textit{n}-dimensional vector between the user and adversary distribution centroids.
However, the adversary only controls the raw biometric measurements (input to the sensor) and furthermore needs to deal with the uncertainty that inevitably occurs with the biometric measurement (i.e., intra-user variation and input noise).
Considering the fact that adversaries cannot alter the DNN, this further restricts the domain of available input manipulations.

\myparagraph{Failed Injections} As adversaries only know one victim sample that is not part of the template, they need to overcome the uncertainty brought by the unknown location of the decision boundary.
In fact, the adversary has no information about the training data and consequently the learned decision boundary (shaded area in Figure~\ref{fig:poisoning-example}).
This means that the adversary does not know whether a crafted poisoning sample would be accepted by the system until he physically carries out the injection (i.e., crafting a poisoning sample, presenting the sample to the system).
The procedure of presenting poisoning samples to the system is an expensive action for the attackers: they might raise suspicion and sample crafting might require resources.
Additionally, the system could also block the template updating procedure after repeated failed attempts.
Therefore, accounting for rejected attempts is fundamental in designing a strategy for the poisoning, as the risk of being detected grows with every attempt.

\section{Attack Method}\label{sec:attack_methodology}

We adopt the idea of facial accessories from Sharif et al.~\cite{Sharif2016} as a building block of our attack: we imagine the adversary can craft coloured glasses and wear them in order to carry out the attack.
The glasses should be re-crafted at each injection step in order to achieve the correct location in features space shown in Figure~\ref{fig:poisoning-example}.
In our attack method, we have to re-define parts of the method in~\cite{Sharif2016} in order to obtain a better resilience against the two challenges mentioned above: input uncertainty and failed injections.

\subsection{Poisoning Sample Generation}\label{sec:adv_sample_gen}

\myparagraph{Formulation}
Formally, given a deep model $f$, a starting sample from the adversary $\vec{x}$, a target sample from the user $\vec{y}$ and their respective feature vectors computed by the model $f(\vec{x})$ and $f(\vec{y})$, we want to find a perturbation on the input features $\delta_x $ that minimizes the following:
\begin{equation}
\arg\min_{\delta_{\vec{x}}} \norm{f(\vec{x} + \delta_{\vec{x}}) - f(\vec{y})}_p,
\end{equation}

where $\norm{\cdot}_p$ is a norm function. 
In order to limit the perturbations into the area of the glasses, we use a binary mask $M_x$ that has the size of the input space and filters out the pixels that do not lie on the glasses frame:
\begin{equation}
\arg\min_{\delta_{\vec{x}}} \norm{f(\vec{x} \cdot (1 - M_x)  + M_x \circ \delta_{\vec{x}}) - f(\vec{y})}_p,
\end{equation}
To account for the glasses smoothness, we look for perturbations that minimize total pixel variation of the added perturbation, computed as a function of each pixel value $p_{i,j}$:
\begin{equation}
V(\delta_{\vec{x}}) = \sum_{k,j}((p_{k,j} - p_{k_+1, j})^2 + (p_{k,j} - p_{k, j+1})^2 )^\frac{1}{2} \: \: \text{\cite{Mahendran2015}}.
\end{equation}
To account for perturbation printability, we account for the non-printability score (NPS) of the adversarial perturbation. Given a set of printable tuples, $P$, NPS is given by:
\begin{equation}
NPS(\delta_{\vec{x}}) = \sum_{p \in \delta_{\vec{x}}} \prod_{p' \in P} | p - p' | \: \: \text{\cite{Sharif2016}}.
\end{equation}
In order to account for the limited injection capabilities and input noise, rather than optimizing individual samples separately we optimize all samples together in a batch.
We then prioritize perturbations that minimize the standard deviation of the samples' distances to the target.
To make the optimization faster, we introduce a weight that prioritizes samples that are further away from the distribution of samples' distance to the target.
So for a set of samples from the adversary $\{\vec{x}_1, ..., \vec{x}_n\}$ and a perturbation $\delta_{\vec{x}}$ the distance $d$ from the mean for sample $i$ is:
\begin{align}
d_i = \abs{\norm{f(\vec{x}_i \cdot (1 - M_{x_i}) + M_{x_i} \circ \delta_{\vec{x}}) - f(\vec{y})}_p - \mu}, \\
\mu = \frac{1}{n}\sum_{i \in n} \norm{f(\vec{x}_i \cdot (1 - M_{x_i}) + M_{x_i} \circ \delta_{\vec{x}}) - f(\vec{y})}_p,
\end{align}
so that the optimization becomes:
\begin{equation}\label{eq:optimization_final}
\begin{split}
\arg\min_{\delta_{\vec{x}}} \sum_{i \in n}d_i\norm{f(\vec{x_i} \cdot (1 - M_{x_i})  + M_{x_i} \circ \delta_{\vec{x}}) - f(\vec{y})}_p \\
+ NPS(\delta_{\vec{x}}) + V(\delta_{\vec{x}}).
\end{split}
\end{equation}
It should be noted that as $M_{x_i}$ changes across different images ($i$) because of the different location of the glasses (depending on eyes landmarks), we refer to $M_{x_i} \circ \delta_{\vec{x}}$ as the operation that fills the positive elements of the mask with the current coloured glasses frame $\delta_{\vec{x}}$.

\myparagraph{Optimization}
\begin{algorithm}[t]
	\caption{\textbf{- Poisoning Samples Generation}: neural network $f$, batch of samples from the adversary  $\{\vec{x}\}_i$, target victim sample $f(\vec{y})$, perturbation regularization $\lambda$, no. of features to update at each step $m$, $\{M_{\vec{x}}\}_i$ masks for glasses location.}\label{alg:generation}
	\begin{algorithmic}[1]
		\INPUT{$f, \{\vec{x}\}_i, f(\vec{y}), \{M_{\vec{x}}\}_i, \lambda, m$}
		\State $\delta_{\vec{x}}^{(1)}$ = \texttt{random\_init()}\Comment{initialize random glasses}
		\State $j \gets 1$
		\For{$(j = 1, j < N, 1)$}
		\State $\{\delta_{\vec{x}}\}_i$ = \texttt{back\_prop($f, \{\vec{x}\}_i, \{M_{\vec{x}}\}_i, \delta_{\vec{x}}^{(j)},  f(\vec{x}_c)$)}
		\State $\Delta \delta_{\vec{x}}^{(j)}$ = \texttt{sel\_top\_f($\{\delta_{\vec{x}}\}_i, m$)}
		\State$\delta_{\vec{x}}^{(j+1)} = \delta_{\vec{x}}^{(j)} + \lambda \Delta \delta_{\vec{x}}^{(j)}.$
		\EndFor
		\For{$(j = N, j < 2N, 1)$}
		\State $\{\delta_{\vec{x}}\}_i$ = \texttt{back\_prop($f, \{\vec{x}\}_i, \{M_{\vec{x}}\}_i, \delta_{\vec{x}}^{(j)}, f(\vec{y})$)}
		\State $\Delta \delta_{\vec{x}}^{(j)}$ = \texttt{sel\_top\_f($\{\delta_{\vec{x}}\}_i, m$)}
		\State$\delta_{\vec{x}}^{(j+1)} = \delta_{\vec{x}}^{(j)} + \lambda \Delta \delta_{\vec{x}}^{(j)}.$
		\EndFor
		\State \textbf{return} $\delta_{\vec{x}}^{(j)} \:\: \forall j \in [N, 2N]$ \Comment{return intermediate results}
	\end{algorithmic}
\end{algorithm}
To find the optimal perturbations, we combine gradient descent with a per-feature iterative approach.
We initialize the glasses frame $\delta_{\vec{x}}^{(0)}$ at random, avoiding too bright or too dark colours and use a Gaussian filter to reduce neighboring pixels variability.
At each step, we first back-propagate the gradient from the network.
Then we weight the backpropagated gradient of sample $x_i$ by $d_i$.
At this point we choose the top $m$ input features (based on the gradient value) which shift the current samples in the desired direction selecting them as those which minimize the total variability and non-printability of the glasses frame, i.e.,  $NPS(\delta_{\vec{x}_i}) + V(\delta_{\vec{x}_i})$.
This way we obtain the changes to apply to the glasses at this step $\Delta \delta_{\vec{x}}^{(j)}$ which we apply to each sample in the batch:
\begin{equation}
\delta_{\vec{x}}^{(j+1)} = \delta_{\vec{x}}^{(j)} + \lambda \Delta \delta_{\vec{x}}^{(j)}.
\end{equation}
We use a regularization parameter $\lambda$ to control the magnitude of pixel changes, to avoid too large steps in the optimization which could lead to sub-optimal adversarial samples.
We report the full procedure in Algorithm~\ref{alg:generation}.

We perform the optimization twice.
After the random initialization, we first optimize the glasses using the adversary's centroid in feature space $x_c$ (Alg.~\ref{alg:generation} Line 4) as target. 
In fact, we want to make sure that at the beginning of the generation of poisoning samples the adversaries wearing the glasses still maintain their original appearance, i.e., the glasses do not affect the location of the attacker's samples in feature space.
After the first optimization has ended, we perform a second optimization where we further modify the glasses so that they now change the adversary appearance onto the victim one, i.e., the glasses shift the location of the attacker's samples in feature space towards the victim.
Algorithm~\ref{alg:generation} returns a list of \textit{intermediate} glasses that correspond to the poisoning samples shown in Figure~\ref{fig:poisoning-example}.
In the next section we use these samples to construct the poisoning attack.

\subsection{Poisoning Attack Injection}

\begin{algorithm}[t]
	\caption{\textbf{- Poisoning Attack}: neural network $f$, victim user template $Y = \{f(\vec{y})\}_i$, known victim sample $\vec{y} \notin Y$, adversary template $\{\vec{x}\}_i$, classifier $C$, perturbation delta $\theta$.}\label{alg:attack}
	\begin{algorithmic}[1]
		\INPUT{$f, \vec{y}, \{\vec{x}\}_i, \lambda, N, C, Y$}
		\State $C$ is trained with $Y$
		\State $\{M_{\vec{x}}\}_i \gets$ \texttt{find\_eyes\_landmarks($\{\vec{x}\}_i$)}
		\State $\{\delta_{\vec{x}}\}^{(j)} \gets \texttt{gen}(f, \{\vec{x}\}_i, f(\vec{y}), \{M_{\vec{x}}\}_i, \lambda, m) $ \label{alg:attack:gen1} \Comment{Alg 1}
		\State $\text{\textsc{iar}} \gets \texttt{count}(C\text{.predict}(\{\vec{x}\}_i) > C\text{.thresh}) $
		\While{\textsc{iar} $< \theta_1$}
		\State $\text{failures} \gets 0$
		\State $j \gets \text{\textsc{heur(..)}}$ \Comment{find perturbations with heuristic}
		\State $\{\vec{x}^*\}_i \gets \{\vec{x} \cdot (1 - M_x) + M_x \circ \delta_{\vec{x}}^{(j)}\}_i$ \Comment{glasses}
		\State $\text{\textsc{iar}}^* \gets \texttt{count}(C\text{.predict}(\{\vec{x}^*\}_i) > C\text{.thresh}) $
		\While {$\text{\textsc{iar}}^* < \theta_2$}
		\State $\text{failures} \gets \text{failures} + 1$
		\State $j \gets j + 1$ \Comment{increase perturbations}
		\State $\{\vec{x}^*\}_i \gets \{ \vec{x} \cdot (1 - M_x) + M_x \circ \delta_{\vec{x}}^{(j)}\}_i$ 
		\State $\text{\textsc{iar}}^* \gets \texttt{count}(C\text{.predict}(\{\vec{x}^*\}_i) > C\text{.thr}) $
		\EndWhile
		\State $Y \gets Y + \{f(\vec{x}^*_k)\}$ \Comment{ s.t. $C\text{.predict}(\vec{x}^*_k) > C\text{.thr} $}
		\State $\text{\textsc{iar}} \gets \texttt{count}(C\text{.predict}(\{\vec{x}\}_i) > C\text{.thr}) $
		\EndWhile
		\State \textbf{return} \texttt{True}
	\end{algorithmic}
\end{algorithm}


After using Algorithm~\ref{alg:generation}, all the necessary intermediate poisoning samples are available for the adversary to use.
However, adversaries need to decide at what point in the optimization process the glasses would lead to an accepted attempt.
Here, we first explain how the poisoning works and then we show how the uncertainty on the decision boundaries can be overcome using population data.

\myparagraph{Poisoning Algorithm}
We report the full algorithm for the poisoning in Algorithm~\ref{alg:attack}.
The algorithm uses Alg.~\ref{alg:generation} to obtain intermediate samples (Alg.~\ref{alg:attack} Line 3).
We use the impostor acceptance rate (IAR) as an indicator for injection (i.e., the proportion of attacker samples that are accepted by the system as legitimate).
We consider that the adversary can successfully inject a sample when at least a $\theta_2$ fraction of the attackers samples (as they are wearing the glasses, $\vec{x}^*$) are accepted by the system (Line 10).
Whenever the adversary attempts to inject a sample, if less than $\theta_2$ of his samples are accepted by the system we consider the attempt a failure.
In this case, the adversary will increase the amount of perturbations on the glasses (move closer to the user's template, see Figure~\ref{fig:poisoning-example}), and attempt again.
In the cases where more than $\theta_2$ samples are accepted, then we consider the attempt successful and inject one of these accepted samples into the current user template (chosen at random, Line 16).
The algorithm stops when at least $\theta_1$ fraction attacker samples while wearing no glasses ($\vec{x}$), is accepted by the system (Line 5).

\myparagraph{Injection Heuristic}\label{sec:heuristic}
In order to minimize the number of attempted injections we develop a \textit{heuristic} to estimate whether a crafted poisoning sample would be accepted given limited information about the training data and classifiers.
The heuristic is based on the intuition that deep models distribute feature space evenly across different users, attempting to separate each user by a similar distance from the others.
Therefore, adversaries can use population data to understand the dynamics of the decision boundaries.
The heuristic is then based on two factors known by the adversary: (i) the number of perturbations applied to the input sample ($j$ in Alg.~\ref{alg:generation}) and (ii) the $L_2$ distance between the poisoning sample and the single known user sample $\norm{f(\vec{x}^*) - f(\vec{y})}_2$.

In practice, the adversary can run the attack for a set of users in the population for which he has knowledge of the template (i.e., excluding the actual victim).
It should be noted that this does not assume white-box knowledge of the model but simply query access (free injection failures).
This poses no significant challenge as user templates can be gathered through social media or just by using publicly available datasets.
In this process, he runs Algorithm~\ref{alg:attack} but replaces Line 7 with an iterative search over the intermediate samples generated by Algorithm~\ref{alg:generation} until he reaches $\theta_2$ accepted samples.
He then collects the number of perturbations applied so far and the $L_2$ distance to the known samples in the template.
With this information, he can find a ``sweet spot'' in the two-dimensional space (i.e., the number of perturbations, distance to samples in the template) that indicates the likelihood of the current poisoning sample $\vec{x}^*$ of being accepted by the template update function. 

Using the data gathered from the population, the adversary now has access to a good estimator of sample acceptance. 
In particular, the closer the current poisoning sample $\vec{x}^*$ is to the center of the two-dimensional distribution, the most likely the sample will be accepted by an unseen template update function.
The adversary therefore uses the heuristic to decide which  $\vec{x}^*$ to inject and falls back to an iterative approach whenever the heuristic fails (as described in the previous paragraph).

\section{Evaluation}
\label{sec:evaluation}
In this section, we first describe the experiment, we then evaluate the attack, show its effect on the error rates and present the results for attack transferability across networks.

\subsection{Experiment Design}\label{sec:experiment-design}

\begin{table}[t]
	\centering
	\begin{tabular}{@{\hskip3pt}c@{\hskip3pt}@{\hskip3pt}c@{\hskip3pt}@{\hskip3pt}c@{\hskip3pt}@{\hskip3pt}c@{\hskip3pt}@{\hskip3pt}c@{\hskip3pt}}
		\toprule
		\textit{Model} & \textit{dataset} & \textit{\# identities} & \textit{output} & \textit{ accuracy} \\
		\midrule
		FaceNet~\cite{Schroff} & \small{VGGFace2} &  8,631 & 512 \tiny{(embed)} & 99.65\% \\\midrule
		VGG16~\cite{Parkhi15} & \small{VGGFace} & 2,622 & 4,096 \tiny{(logits)} & 98.95\% \\\midrule
		ResNet-50~\cite{cao2018vggface2} & \small{VGGFace2} & 8,631 & 2,048 \tiny{(logits)} & $>$99\% \\\bottomrule
	\end{tabular}
	\caption{
		Info for models and datasets used. The \textit{output} column reports the dimensionality and type of the layer we used to extract facial features. Accuracy refers to the accuracy computed on the labeled faces in the wild dataset~\cite{Huang2007} (LFW), and the figures are taken from the respective papers.
	}
	\label{tab:considered_dnns}
\end{table}

\myparagraph{Classifiers} we decide to consider three different classifiers: (i) \texttt{centroid}~\cite{Biggio2013}, (ii) \texttt{maximum} ~\cite{Biggio2013} and (iii) one-class Support Vector Machine (\texttt{SVM}). 
The \texttt{centroid} classifier computes the distance between the new sample and the centroid of the training data.
The \texttt{maximum} classifier computes the distance between the new sample and the closest training sample.
Both classifiers use $L_2$ distance to compute distances between samples; we tried using $L_1$ and we obtained similar results.
All three classifiers perform an authentication decision comparing the computed distance to a pre-set threshold.
We decide to use a linear kernel for \texttt{SVM} as this was the best performing one in terms of recognition performance.
To account for the age of samples in the template, i.e., older samples might be considered less relevant than recent samples, we include two different weighting schemes: (i) \textit{flat}, (ii) \textit{sigmoid}.
In the flat weighting schemes, each sample in the template is considered equally important in the decision, while in the sigmoid scheme the weight of each sample $w_i$ is computed with a sigmoid function prioritizing recent samples: $w_i = \frac{1}{(1+e^{-x_i})}$,
where $x_i \in [-5, 5]$ indicates how recent the sample is (the lower the older the sample is).
Intuitively, prioritizing recent samples in the decision will make the poisoning attack faster.

\myparagraph{Update Policy} For the template update procedure, we choose to use self-update with \textit{infinite window update} policy~\cite{Biggio2012, Kloft2010}.
In other words, whenever a new sample is accepted by the classifier it is subsequently added to the training set.
This choice is similar to the unsupervised update process used by Apple's FaceID or Siri, where newly obtained samples are added to the template if the sample quality (i.e., low level of noise) is sufficient~\cite{faceidsecurityguide, personalized-hey-siri}.

\myparagraph{Considered DNN}
We consider three different state-of-the-art convolutional network architectures: FaceNet~\cite{Schroff}, the VGG16 Face descriptor~\cite{Parkhi15} and ResNet-50~\cite{cao2018vggface2}.
FaceNet uses the Inception-ResNet-v1 architecture from~\cite{szegedy2017inception}, and is trained with triplet loss to produce 512-dimensional embeddings.
The VGG-Face descriptor is based on the VGG16 architecture~\cite{simonyan2014very} and has been trained on the VGGFace dataset, which contains 2,622 identities.
Both ResNet-50 and FaceNet are trained on the newer VGGFace2~\cite{cao2018vggface2} dataset which contains 8,631 different identities with on average over 340 images per individual.
For VGG16 and ResNet-50 we use the $L_2$-normalized \textit{logits} layer as output rather than training the models with triplet loss.
The details of the models and datasets are reported in Table~\ref{tab:considered_dnns}.

\myparagraph{Input Preprocessing}
All images are first aligned using the landmark based model in~\cite{7553523}.
All the remaining preprocessing follows the models' guidelines: FaceNet inputs are pre-whitened, VGG16 and ResNet-50 inputs use fixed image standardization.
We don't use any random cropping or flipping for the images.
The images dimensions in input are 160x160 for FaceNet and 224x224 for the remaining models. 
When testing images across different models we resize them to the correct size using bilinear interpolation.

\myparagraph{Experiment Description}
\begin{table}[tp]
	\centering
	\begin{tabular}{ccc|cc|cc}
		\toprule
		& \multicolumn{2}{c}{\texttt{centroid}} & \multicolumn{2}{c}{\texttt{maximum}} &\multicolumn{2}{c}{\texttt{SVM}}\\
		\textit{Model} & \textit{\textsc{eer}} & \textit{thr.} & \textit{\textsc{eer}} & \textit{thr.} & \textit{\textsc{eer}} & \textit{thr.} \\
		\midrule
		FaceNet & 0.9\% & 0.959 & 1.3\% & 0.901 & 1.0\% & 0.121\\\midrule
		VGG16 & 2.2\% & 0.590 & 3.0\% & 0.564 & 1.9\% & 0.041\\\midrule
		ResNet-50 & 1.3\% & 0.923 & 1.9\% & 0.876 & 1.7\% & 0.098\\\bottomrule
	\end{tabular}
	\caption{
		Performance of the face-recognition models in terms of EER.
		The classifiers thresholds are set at the EER on a separate subset of 100 users that are not used in the attack evaluation.
	}
	\label{tab:eer}
	\vspace{-0.3cm}
\end{table}

All the experiments make use of the testing part of the VGGFace2 dataset, which contains 500 separate users. 
Before we evaluate the attack's effectiveness, we first set up the system as follows.
First, we randomly choose 100 users to compute the classifiers' thresholds.
For each user we use 10 samples for training and test against all the remaining user samples and a randomly selected sample of different users ($>$100,000 authentication attempts are used to find the EER threshold).
The results are reported in Table~\ref{tab:eer}.
Afterwards, we split the 400 remaining users into two equally sized groups and we treat the two groups as adversaries and victims separately.
We randomly choose 1,000 attacker-victim pairs for the evaluation. 
These chosen pairs are the same across each model considered. 

\begin{figure}[t]
	\centering
	\includegraphics[width=\columnwidth]{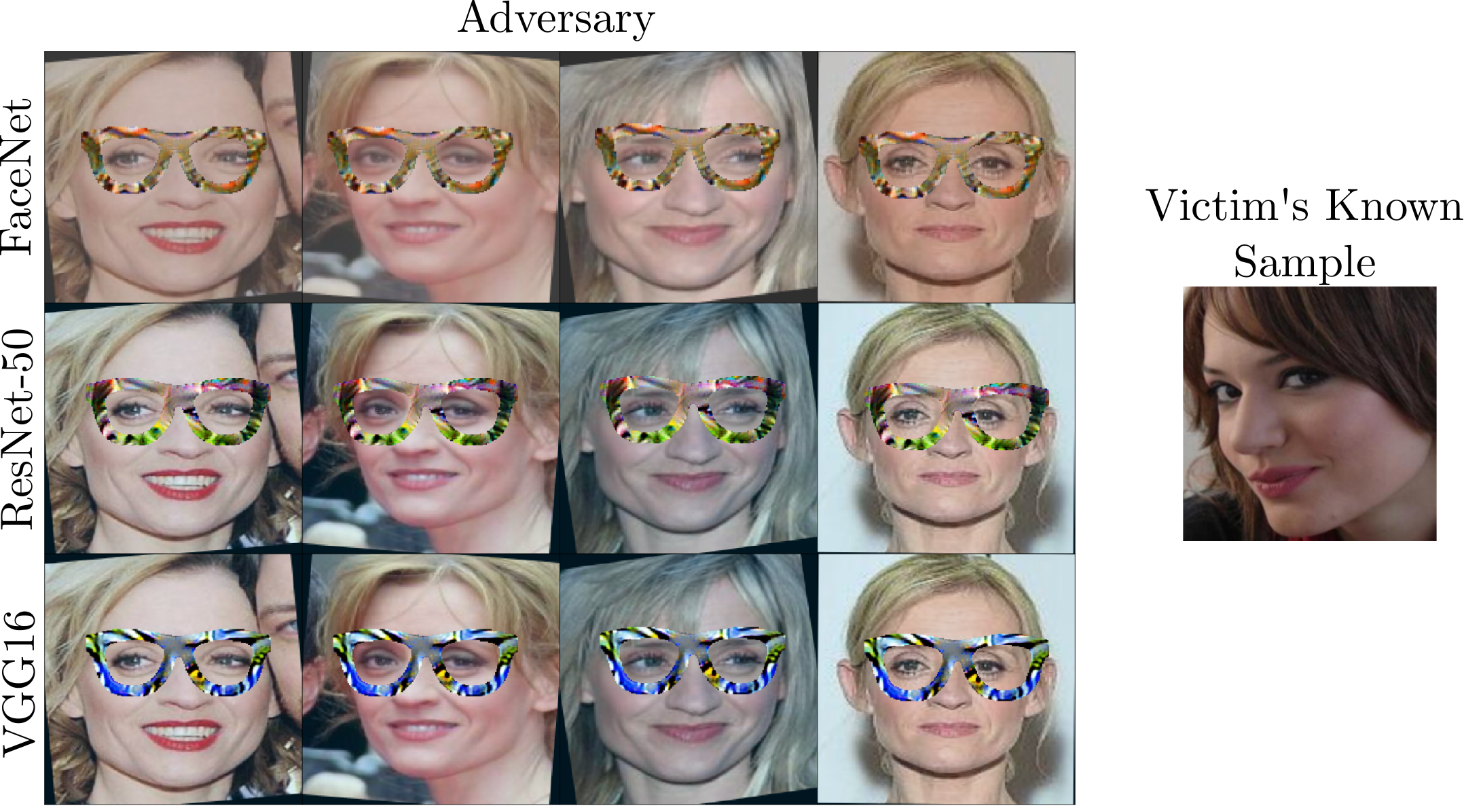}
	\caption{Poisoning samples for an adversary-victim pair. The figure shows how the \textit{same} glasses can be applied to the user across different intra-user variations (e.g., pose). 
	}
	\label{fig:faces}
\end{figure}

\myparagraph{Attack Implementation}
We always use Algorithm~\ref{alg:attack} for the evaluation.
We define the attack as successful when at least half of the adversary template is accepted by the classifier and we use the same to compute the injection success rate, i.e., $\theta_1 = \theta_2 = 0.5$.
As mentioned in Section~\ref{sec:systemandthreatmodel}, the update and authentication thresholds are identical (found in Table~\ref{tab:eer}), so that when a sample passes authentication it is automatically added to the template.
The heuristic of Section~\ref{sec:heuristic} is computed using only 10 randomly chosen attacker-victim pairs.
The glasses are positioned based on the eyes location (computed using a face landmark extractor) and occupy  on average 8.59\% of the total pixels in input.
Figure~\ref{fig:faces} shows a set of poisoning samples for an attacker and the relative victim.
We only use front facing images from the adversary (up to 50 images), as adversaries have control over their pose as they carry out the attack, and do not need to optimize for their own intra-user variation when attempting the attack.
At the moment of injection, the injected sample is chosen at random from the currently accepted samples.
We use $\norm{\cdot}_2$ for the optimization of Equation~\ref{eq:optimization_final} as $L_2$ is used by the classifiers for the decision.
The parameter $\lambda$ is set to 4, which indicates that the maximum change for a pixel in an iteration is 4 out of a range of $[0, 255]$\footnote{code available at \url{https://github.com/ssloxford/biometric-backdoors}}.

\begin{figure*}[t]
	\centering
	\includegraphics[width=\textwidth,height=9cm]{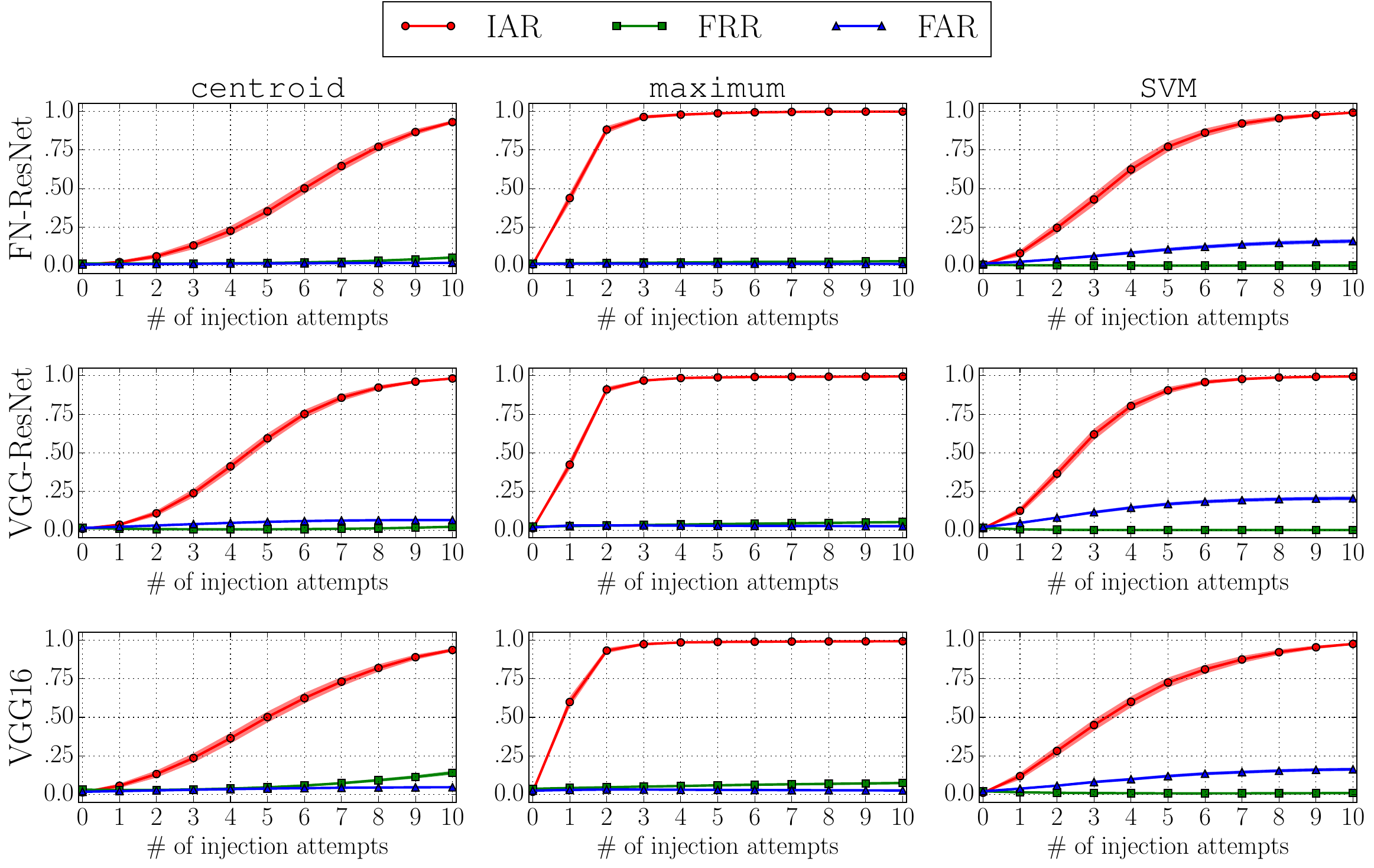}
	\caption{IAR, FAR and FRR changes over the course of the poisoning attack. Shaded areas show the standard deviation over each adversary-victim pair in the dataset. The shown results are for the \texttt{centroid} classifier with flat weights.}
	\label{fig:rates}
\end{figure*}

\subsection{Attack Success and Effect on Error Rates} \label{sec:effect_on_error_rates}

\begin{table*}[t]
\centering
\begin{tabular}{c@{\hskip3pt}c@{\hskip3pt}@{\hskip3pt}c@{\hskip3pt}@{\hskip3pt}c@{\hskip3pt}|@{\hskip3pt}c@{\hskip3pt}@{\hskip3pt}c@{\hskip3pt}@{\hskip3pt}c@{\hskip3pt}|@{\hskip3pt}c@{\hskip3pt}@{\hskip3pt}c@{\hskip3pt}@{\hskip3pt}c@{\hskip3pt}|@{\hskip3pt}c@{\hskip3pt}@{\hskip3pt}c@{\hskip3pt}@{\hskip3pt}c@{\hskip3pt}|@{\hskip3pt}c@{\hskip3pt}@{\hskip3pt}c@{\hskip3pt}@{\hskip3pt}c@{\hskip3pt}|@{\hskip3pt}c@{\hskip3pt}@{\hskip3pt}c@{\hskip3pt}@{\hskip3pt}c@{\hskip3pt}}
\toprule
& \multicolumn{6}{c}{\texttt{centroid}} & \multicolumn{6}{c}{\texttt{maximum}} &\multicolumn{6}{c}{\texttt{SVM}}\\
\cmidrule(l){2-7}\cmidrule(l){8-13}\cmidrule(l){14-19}
& \multicolumn{3}{c}{\textit{flat}} & \multicolumn{3}{c}{\textit{sigmoid}} & \multicolumn{3}{c}{\textit{flat}} & \multicolumn{3}{c}{\textit{sigmoid}}& \multicolumn{3}{c}{\textit{flat}} & \multicolumn{3}{c}{\textit{sigmoid}} \\\textit{Model}&\textit{1\textsuperscript{st}} & \textit{3\textsuperscript{rd}} & \textit{10\textsuperscript{th}} & \textit{1\textsuperscript{st}} & \textit{3\textsuperscript{rd}} & \textit{10\textsuperscript{th}} & \textit{1\textsuperscript{st}} & \textit{3\textsuperscript{rd}} & \textit{10\textsuperscript{th}} & \textit{1\textsuperscript{st}} & \textit{3\textsuperscript{rd}} & \textit{10\textsuperscript{th}} & \textit{1\textsuperscript{st}} & \textit{3\textsuperscript{rd}} & \textit{10\textsuperscript{th}} & \textit{1\textsuperscript{st}} & \textit{3\textsuperscript{rd}} & \textit{10\textsuperscript{th}} \\\midrule
FN-ResNet  & 1\% & 9\% & 77\% & 6\% & 55\% & 95\% & 40\% & 93\% & 95\% & 86\% & 97\% & 97\% & 6\% & 35\% & 81\% & 10\% & 49\% & 73\%\\
\midrule
VGG16  & 4\% & 19\% & 85\% & 12\% & 66\% & 100\% & 61\% & 98\% & 100\% & 92\% & 100\% & 100\% & 9\% & 42\% & 94\% & 12\% & 50\% & 71\%\\
\midrule
VGG-ResNet  & 2\% & 20\% & 97\% & 10\% & 86\% & 99\% & 38\% & 95\% & 96\% & 98\% & 99\% & 99\% & 9\% & 54\% & 87\% & 13\% & 63\% & 80\%\\
\bottomrule
\end{tabular}
\caption{
Attack success rates for each considered model and classifier.
Success is defined as $>$50\% IAR after $i$ samples injections (either one, three or ten).
Each figure is calculated on the (same) 1000 randomly chosen pairs of attacker-victim.
}
\label{tab:big_result}
\end{table*}

Since the attack is generally successful in the long-term (as more malicious samples enter the template), we focus the analysis on the number of injection attempts required and the effect on the recognition rates as the poisoning progresses.
For a report on the attack success rate as defined in Section~\ref{sec:attack_methodology}, see Table~\ref{tab:big_result}.
We found that the heuristic presented in Section~\ref{sec:heuristic} can compensate for the limited decision boundaries knowledge of the attacker: every time the attacker used the heuristic to craft an intermediate sample to inject, the sample was accepted in more than 79\% of cases (79\%, 83\% and 82\% for FaceNet, VGG16 and ResNet-50, respectively).
The results show the performance of the attacker \textit{including} failed injections.

We monitor three different rates: (i) false accept rate (FAR), (ii) false reject rate (FRR), (iii) impostor accept rate (IAR). 
For an attacker-victim pair, FAR is computed as the proportion of samples belonging to ``other users'' (users that are not the attacker or the victim) that are accepted by the classifier.
Similarly, FRR is the proportion of victim's samples that are rejected by the classifier, excluding the 10 samples in the training set.
IAR is defined as the proportion of samples belonging to the adversary that is accepted by the classifier without wearing glasses.
Ideally, the attack should make sure that IAR increases as fast as possible while not changing FRR and FAR.
Increases in FRR are suspicious because the legitimate user cannot authenticate smoothly anymore (lots of rejections).
Consequently, users might contact the system administrator to report the problem or switch to more convenient authentication factors (when other options are available).
Increases in FAR are less suspicious,  depending on how often other users attempt impersonation attacks with their own biometric trait.

Figure~\ref{fig:rates} shows how the above-mentioned rates vary during the poisoning attack, for the flat weighting scheme.
The three models respond similarly to the poisoning, while different classifiers show different error rate changes.
Comparing the classifiers with each other shows that \texttt{maximum} classifier is particularly vulnerable to this attack: two injected samples lead to IAR $>$90\% on average, only with marginal increments in the success rate for additional injections.

Overall the changes in FRR and FAR are minimal.
A small FAR increase can be seen for the \texttt{SVM} classifier.
This might be due to the linear kernel function used for the classifier, which cannot fit  the training data well when poisoning samples are added 
and therefore includes wider areas of the feature space.
The minimal changes in FRR are in contrast with what has been shown in previous work~\cite{Biggio2013}, where FRR (referred to as genuine accept rate) quickly increased to values above 40\% after 5 injected samples and over 90\% over 10 injected samples.
This is mainly due to a combination of the initial enrolment data size and the type of window used for the self-update procedure.
As noted previously~\cite{Kloft2010}, fewer samples in the training data lead to faster poisoning (ten in our case, five in~\cite{Biggio2013}) and the choice of discarding older samples~\cite{Biggio2013} with the finite window policy leads to the user being removed from the template.
As mentioned in Section~\ref{sec:experiment-design}, it should be noted that modern systems tend to reflect the design choices in this work, both for speaker and face recognition~\cite{personalized-hey-siri,faceidsecurityguide}).
A more detailed and long-term analysis is required to investigate the trade-offs between these choices and the performance of the system in this scenario, which we leave for future work. 

\subsection{Transferability}\label{sec:transferability}

\begin{table*}[t]
	\centering
	\begin{tabular}{@{\hskip3pt}c@{\hskip3pt}@{\hskip3pt}c@{\hskip3pt}@{\hskip3pt}c@{\hskip3pt}@{\hskip3pt}c@{\hskip3pt}|@{\hskip3pt}c@{\hskip3pt}@{\hskip3pt}c@{\hskip3pt}@{\hskip3pt}c@{\hskip3pt}|@{\hskip3pt}c@{\hskip3pt}@{\hskip3pt}c@{\hskip3pt}@{\hskip3pt}c@{\hskip3pt}|@{\hskip3pt}c@{\hskip3pt}@{\hskip3pt}c@{\hskip3pt}@{\hskip3pt}c@{\hskip3pt}}
		\toprule
		& \multicolumn{3}{c}{FaceNet} & \multicolumn{3}{c}{FaceNet-\footnotesize{CASIA}} & \multicolumn{3}{c}{ResNet-50} & \multicolumn{3}{c}{VGG16}
		\\
		\textit{Model} & \texttt{centroid} & \texttt{maximum} & \texttt{SVM}& \texttt{centroid} & \texttt{maximum} & \texttt{SVM}& \texttt{centroid} & \texttt{maximum} & \texttt{SVM}& \texttt{centroid} & \texttt{maximum} & \texttt{SVM} \\\midrule
		FaceNet  & \textbf{77}\%  & \textbf{95}\%  & \textbf{81}\%  & 22\%  & 27\%  & 24\%  & 12\%  & 12\%  & 14\%  & 9\%  & 11\%  & 9\% \\
		FaceNet-\footnotesize{CASIA}  & 9\%  & 14\%  & 10\%  & \textbf{85}\%  & \textbf{99}\%  & \textbf{88}\%  & 11\%  & 11\%  & 12\%  & 9\%  & 10\%  & 9\% \\
		ResNet-50  & 13\%  & 20\%  & 14\%  & 23\%  & 30\%  & 27\%  & \textbf{97}\%  & \textbf{96}\%  & \textbf{87}\%  & 13\%  & 16\%  & 13\% \\
		VGG16  & 5\%  & 8\%  & 6\%  & 11\%  & 16\%  & 16\%  & 14\%  & 17\%  & 18\%  & \textbf{85}\%  & \textbf{100}\%  & \textbf{94}\% \\
		\bottomrule
	\end{tabular}
	\caption{
		Transferability results of the poisoning attack across different models.
		The reported figures are the success rates of the attack uniquely using information from the source model (source is on the rows, target is on the columns).
		Bold values refer to same-model success rates (also found in Table~
\ref{tab:big_result}).
The rates correspond to the success at the 10\textsuperscript{th} injection attempt.
The heuristic is always fit on the target system (i.e., both target model and classifier).
}
\label{tab:transferability}
\end{table*}

We analyze the transferability of the attack across the models, reporting results in Table~\ref{tab:transferability}.
In addition to the previously considered models, we add FaceNet-\textsc{casia} which has the same architecture as FaceNet, but is trained on the CASIA Webface dataset~\cite{yi2014learning} instead of the VGGFace2 dataset.
This allows us to investigate the influence of the training data on the transferability.
Each attack repetition uses the source model as a surrogate: using the source to compute the poisoning samples 
(Alg.~\ref{alg:attack} Line 3),
but using the target model to compute the sample acceptances
(Alg.~\ref{alg:attack} Line 4, 9 and 14).

Table~\ref{tab:transferability} reports the success rate, defined as $>$50\%  IAR after 10 injected samples, for source-target model pairs.
We found that the low success rate is mostly caused by the fact that the optimization of Algorithm~\ref{alg:generation} did not lead to accepted samples (85\% of attacker-victim pairs) rather than not enough samples being accepted after 10 injections (2\%).
Comparing the white-box results with the across models ones, we see that in most cases there is a tenfold decrement in the success rates, which could still make the attack viable in black-box scenarios.
We don't find particular differences across classifiers, with the \texttt{maximum} being slightly weaker against this attack.
These findings are in line Sharif et al.~\cite{Sharif2017} who report similar transferability for dodging attacks across different networks.

Table~\ref{tab:transferability} does not show any particularly evident trend in the transferability across architectures, showing that some transferability applies for each pair of models.
Additionally, while it would seem that higher EER would lead to higher chance of attacks, 
we find that baseline EER and success rate are only weakly correlated ($r$=-0.18).
This suggests that transferability properties rely on less intuitive combinations of both training datasets and architectures.
It should be noted that the success rate of attacks across different models could be improved, e.g., by using optimizations that are not tailored to an individual networks~\cite{Sharif2017} or by obtaining better approximations of the user template.

\section{Poisoning Countermeasures}\label{sec:poisoning_detection}

Here, we discuss possible countermeasures and propose and evaluate a new detection method based on angular similarity.

\subsection{Detecting Adversarial Samples}

One indirect approach to limit the feasibility of a poisoning attack is to increase the difficulty of crafting adversarial samples with the required properties.
As an example, in adversarial training~\cite{Papernot2016,carlini2017towards,goodfellow2014explaining}, the training data includes adversarial samples specifically crafted to increase the model resilience against them at test time.
With gradient masking, the network gradient is hidden during training, making the network robust to small perturbations in the input~\cite{papernotsok2018, Lyu2016, gu2014towards}.
In AI\textsuperscript{2}~\cite{Gehr2018} the approach aims to approximate the functions learnt by neural networks with abstract primitives, where security guarantees can be certified with sounder methods.
Similarly, when an attack vector is identified, such as wearing coloured glasses, measures that are specific to the detection of such attack can be implemented (e.g., detecting whether the pixels in input present anomalous sharpness).

\subsection{Template Anomalies}

Analyzing anomalies in the template is another approach to stop poisoning attacks, where the goal is to detect whether sets of samples that are added to the template are anomalous.
Biggio et al.~\cite{biggio2015adversarial} propose a technique named \textit{template sanitization} which consists in defining a sanitization hypersphere around the current template distribution.
Whenever series of $k$ consecutive updates quickly drift the current user centroid outside of the current hypersphere, then such updates are discarded and the previous centroid is restored.
The underlying assumption is that genuine template updates exhibit a less biased and more random behavior.
A theoretical analysis of the security of online template updates is given by Kloft et al.~\cite{Kloft2010}, where a protection mechanism based on monitoring the false positive rate is introduced.
Such method suggests that the system could monitor the number of false rejections using a \textit{hold-out} dataset to check whether these exceed a fixed thresholds.
In other words, when the users legitimate samples start to be rejected as the poisoning progresses, updates might be discarded and the centroid reset to a previous ``safe'' state.

\myparagraph{Detection Trade-off}
Both works~\cite{Kloft2010,biggio2015adversarial} show that such countermeasures can successfully thwart the progress of a poisoning attack.
The underlying assumptions of the methods is that poisoning samples present a behavior that is not shown by legitimate updates: either quickly drift the centroid outside of a set hypersphere~\cite{biggio2015adversarial} or increase the number of false rejections~\cite{Kloft2010}.
However, neither study offers an analysis of the trade-off between poisoning detection and legitimate template update acceptance i.e., legitimate samples might show the behaviour that is labeled as anomalous.

\begin{figure}[t]
	\centering
	\begin{subfigure}[b]{0.425\textwidth}
		\centering
		\includegraphics[width=\textwidth,height=2cm]{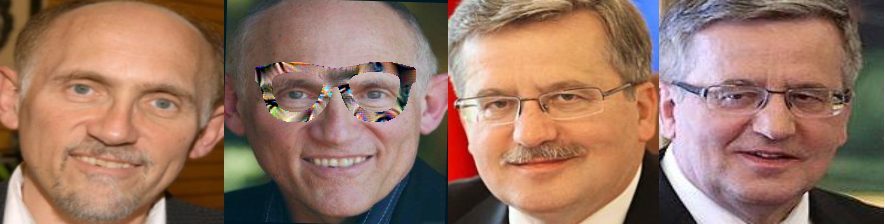}
		\caption%
		{{\small from the left, adversary, adversary during the poisoning attack, victim with facial hair and without facial hair.}}    
		\label{fig:cm_faces}
	\end{subfigure}
	\hfill
	\begin{subfigure}[b]{0.45\textwidth}  
		\centering 
		\includegraphics[width=\textwidth,height=3cm]{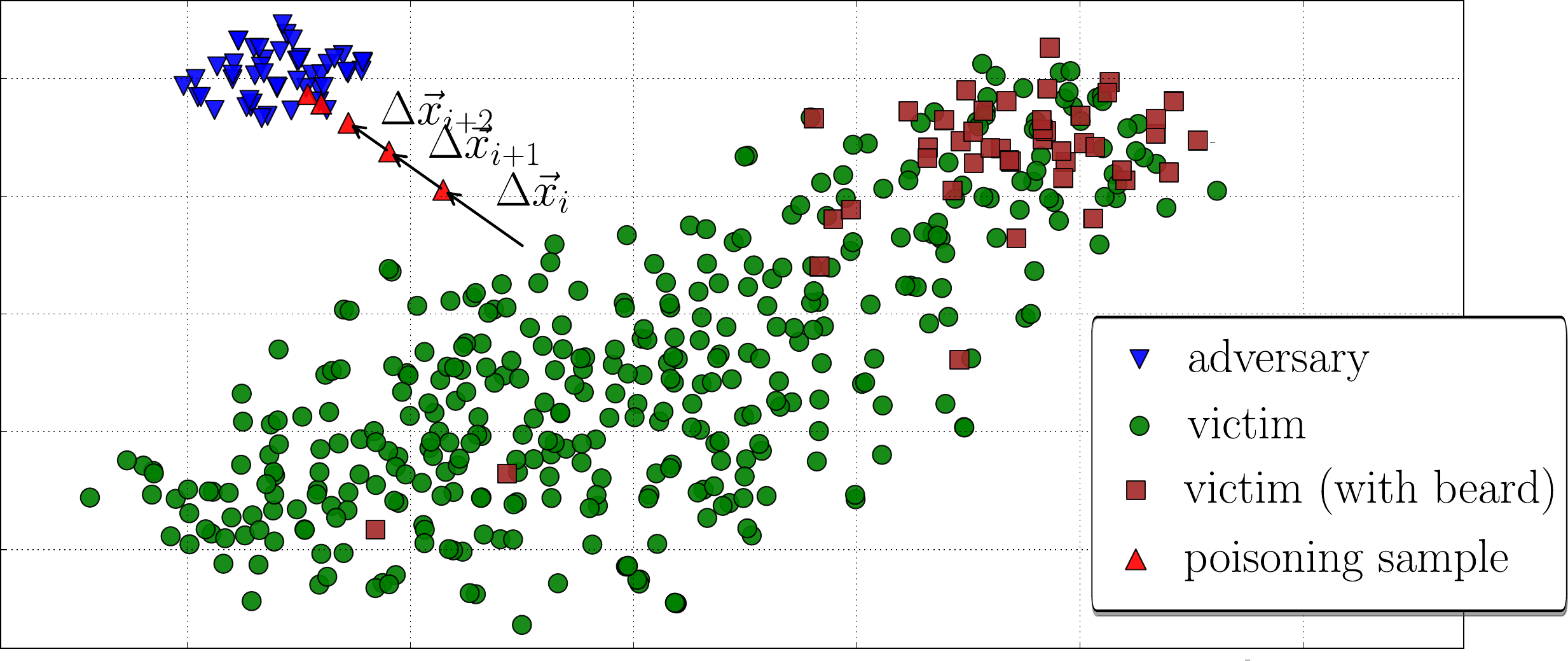}
		\caption[]%
		{{\small embedding space visualized.}}    
		\label{fig:cm_explained}
	\end{subfigure}
	\vskip\baselineskip
	\caption{Poisoning detection. The figure shows victim's, adversary's and poisoning samples. Samples' locations are computed by chaining PCA and t-SNE reduction. Figure~\ref{fig:cm_explained} that consecutive poisoning samples have consistent angular similarity with one another when compared with the victim centroid. Similarly, other intra-user variation factors (e.g., beards) might reside in a specific direction when compared with the center of the victim distribution.}
\end{figure}

For~\cite{Kloft2010}, we have shown that false rejections in our scenario present only marginal increases over the course of the poisoning, and the method additionally assumes the availability of an hold-out dataset for the user, which is not available in a biometric authentication scenario.
For~\cite{biggio2015adversarial}, we notice that several intra-user variation factors that might affect the location of the sample in feature space in a predictable and consistent way, such as: facial hair, pose, age.
Figure~\ref{fig:cm_explained} shows an example of such behaviour: samples where the user has facial hair (i.e., a moustache, see Figure~\ref{fig:cm_faces}) cluster together in a specific region of feature space.
Such a region may be separate from the known user template, meaning that limiting the template updates to a set hypersphere would stop legitimate updates in this scenario~\cite{biggio2015adversarial}.
In the following, we present a new poisoning countermeasure and evaluate its detection trade-off with legitimate samples.

\begin{table*}[t]
	\centering
	\begin{tabular}{@{\hskip2pt}c@{\hskip2pt}c|c|c|cc|cc|cc|cc}
		\toprule
		& \multicolumn{1}{c}{FaceNet} & \multicolumn{1}{c}{ResNet-50} & \multicolumn{1}{c}{VGG16} &\multicolumn{2}{c}{{\footnotesize FaceNet$\,\to\,$ResNet-50}} &\multicolumn{2}{c}{{\footnotesize ResNet-50$\,\to\,$ FaceNet}} &\multicolumn{2}{c}{{\footnotesize FaceNet$\,\to\,$VGG16}} &\multicolumn{2}{c}{{\footnotesize VGG16$\,\to\,$ResNet-50}} \\\textit{factor} & \textsc{eer} & \textsc{eer} & \textsc{eer} &\textsc{far} & \textsc{frr} & \textsc{far} & \textsc{frr} & \textsc{far} & \textsc{frr} & \textsc{far} & \textsc{frr} \\\midrule
		age & 6.8\% & 7.1\% & 4.5\% & 7.4\% & 6.8\% & 6.6\% & 7.2\% & 7.0\% & 2.2\% & 4.8\% & 12.0\% \\glasses & 14.3\% & 14.6\% & 11.3\% & 14.5\% & 14.6\% & 14.3\% & 14.3\% & 18.7\% & 5.8\% & 7.7\% & 27.0\% \\facial hair & 7.6\% & 7.1\% & 4.8\% & 6.8\% & 7.5\% & 8.0\% & 7.2\% & 6.9\% & 2.6\% & 4.6\% & 12.6\% \\pose & 6.3\% & 6.2\% & 4.0\% & 6.2\% & 6.1\% & 6.3\% & 6.3\% & 6.5\% & 1.9\% & 4.0\% & 11.1\% \\\textbf{all} & \textbf{7.7\%} & \textbf{7.9\%} & \textbf{5.4\%} & \textbf{7.7\%} & \textbf{7.6\%} & \textbf{7.9\%} & \textbf{7.9\%} & \textbf{7.7\%} & \textbf{2.6\%} & \textbf{5.4\%} & \textbf{13.9\%} \\\bottomrule
	\end{tabular}
	\caption{
		Rates for the poisoning detection. The top row indicates which model is considered for the evaluation: $a\,\to\,b$ indicates that the detection is trained on $a$ and applied to $b$.
		The values are computed on the \texttt{centroid} with flat weights.
		For same-model cases we report EER. For across-model cases we report the FAR and FRR for the detection on the target model obtained after choosing
		the threshold at EER on the source model.
		The ``all'' row is computed using all the updates from each other factor.}
	\label{tab:countermeasure}
\end{table*}

\subsection{Angular Similarity Detection}
We propose a new detection technique, based on the rationale that poisoning samples will all lie in a predetermined direction in feature space with respect to the current legitimate user centroid.
The direction is determined based on the location of the attacker samples.
This is visualized in Figure~\ref{fig:cm_explained}.
Given the user's current centroid $\vec{x}_c$ and a set of template updates $\{\vec{x}_i, \vec{x}_{i+1}, ..., \vec{x}_{i + n} \}$, which we refer to as an \textit{update sequence}, we compute the direction of the update at time $i$: $\Delta\vec{x}_{i} = \vec{x}_c - \vec{x}_i,$
and we can obtain the directions at each step as $\{\Delta \vec{x}_i, ..., \Delta \vec{x}_{i+n}\}$.
We then compute the angular similarity for pairs of consecutive updates with the cosine similarity as:
\begin{equation}
\label{eq:cm}
\cos \theta_{i+1}= \frac{\Delta \vec{x}_{i} \cdot \Delta \vec{x}_{i+1}}{\norm{\Delta \vec{x}_{i}} \norm{\Delta \vec{x}_{i+1}}}.
\end{equation}
The underlying intuition is that $\cos \theta_{i}$ will be higher for pairs of poisoning samples compared to legitimate updates because the attack needs to shift user centroid towards the adversary's, which lies in a specific pre-defined direction, see Figure~\ref{fig:cm_explained} for reference.

\subsection{Detection Evaluation}

\myparagraph{Setup}
We use the Google Vision API\footnote{https://cloud.google.com/vision/} to extract attributes for four intra-user variation factors: pose, facial hair and (sun)glasses.
Since the API does not return a value for age, we use the age-estimator in~\cite{yang2018ssr}.
For pose we only consider the \textit{pan} angle as a factor, that is the horizontal angle of the face (a pan of 90\degree{} corresponds to a face looking sideways).
For age we group samples into subgroups of samples with the same age, using ranges of three years for each subgroup.
This way, for each of the 1,000 attacker-victim pairs considered in the previous section, we consider sequences of updates using the samples in the victim's testset as follows:
\begin{itemize}[leftmargin=0.6cm]
	\item \textbf{age}: update sequences are in 3-years age span;
	\item \textbf{pose}: we create two update sequences by choosing samples with pan angle $\geq 30$\degree{} or samples with  pan angle $\le 30$\degree{};
	\item \textbf{facial hair}: we create an update sequence where the user's samples have facial hair;
	\item \textbf{glasses}: we create an update sequence where the user is wearing glasses or sunglasses;
	\item \textbf{poisoning}: we consider sequences of poisoning samples obtained from carrying out the attack of Sec.~\ref{sec:evaluation}.
\end{itemize}
We choose a random order for the samples in these sequences, excluding the poisoning one, and we evaluate the angular similarity of consecutive samples with Equation~\ref{eq:cm}.
We consider 1,132,994 legitimate updates, distributed across the attacker-victim pairs.
By comparing legitimate and poisoning sequences we choose a threshold for the detection of too similar subsequent updates.
We then show how IAR changes during the poisoning attack with detection in place.
We present the results of our analysis on the \texttt{centroid} classifier with flat weights; other classifiers performed similarly.

\begin{figure}[t]
	\centering
	\includegraphics[width=\columnwidth,height=3cm]{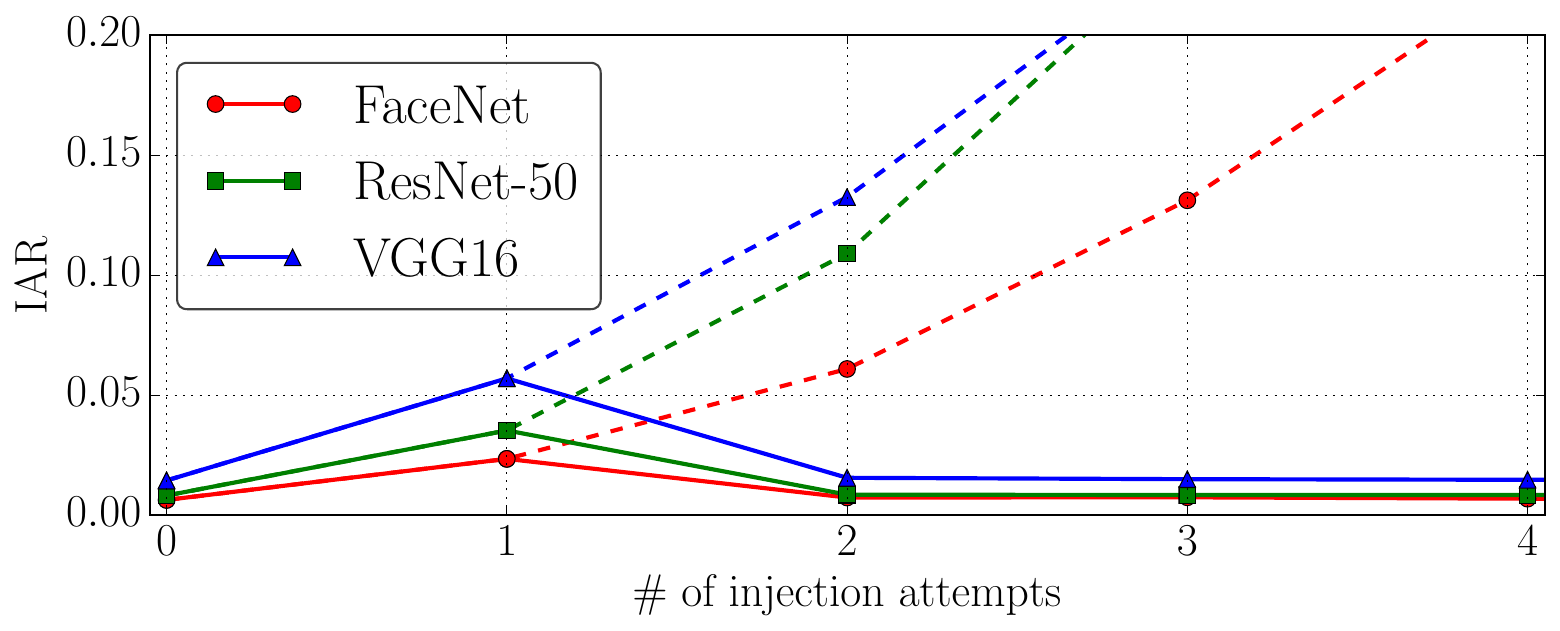}
	\caption{Poisoning detection. The plot shows the IAR comparing when the poisoning detection is in place (solid lines) and when it is not (dashed lines), for the \texttt{centroid} classifier with flat weights and the different models (same-model scenarios). The angular similarity threshold is set at EER on all the samples in the considered update sequences.}
	\label{fig:countermeasure_rates}
\end{figure}

\myparagraph{Results}
We report in Table~\ref{tab:countermeasure} the results for the poisoning detection for each of the considered factors.
Each row reports the results for a specific factor, that is, all pairs in the update sequences of that factor are evaluated against the pairs in the poisoning sequences; the last \textit{all} row reports the result of using all the measurements together.
We see that the detection performance in terms of EER lies around 6-7\% for most factors and models (VGG16 model performs better), with the exception of glasses, which create an increase in the EER up to 14\%.
We find that glasses, in particular sunglasses, create a very predictable and consistent effect on the location of samples in feature space, leading to greater angular similarity of consecutive updates.
This can be explained with sunglasses being less frequent in the training data compared to the other factors such as pose and age, that the network learns how to ignore during optimization.

We test the transferability of the detection by applying the threshold learned on one model to different models.
The last four columns of Table~\ref{tab:countermeasure} show that this is the case: 
the obtained FAR and FRR on the target model closely resemble the EER obtained in the source model, excluding VGG16 where FRR is slightly higher.

We report in Figure~\ref{fig:countermeasure_rates} the IAR over the course of the poisoning with and without detection in place, setting the threshold at EER for all the considered update sequences.
We find that $>$99\% of attacks are detected at the second injection attempt, when the IAR is still low for the attacker to authenticate consistently ($<$6\%).
When consecutive poisoning samples are detected, the system removes the malicious samples from the template, resetting the  baseline IAR.

\section{Discussion}\label{sec:discussion}

In this section we discuss future research directions, including how to enhance the attack practicality and the attack transferability in black-box scenarios.

\myparagraph{Attack Practicality}
Using colored glasses has been shown to be a practical way of fooling face authentication~\cite{Sharif2016}.
However our poisoning attack requires greater control over the embedding space compared to simple impersonation as the exact distance between samples in embedding space influences the attack outcome.
Adversaries could improve such control by accounting for as many factors of intra-user variation as possible during the optimization (Algorithm~\ref{alg:generation}), such as pose, lighting, facial accessories and backgrounds.
In other words, adversaries would build a training dataset which captures all factors of variations of their own face appearance and optimize the glasses by using all the samples in the dataset.
Correctly capturing these factors would lead to better handling of the input uncertainty at the time of injection and consequently to a better control over the injected samples.
Adversaries could also obtain ``offline'' insights on whether a pair of glasses generalizes well against the various factors that cause input uncertainty as follows: split the above mentioned dataset into a training and testing part, use the training part for Algorithm~\ref{alg:generation} but using the testing part to simulate the actual injection in Algorithm~\ref{alg:attack}.
This would allow the adversary to refine and evaluate the poisoning samples before the time of attack to minimize the risk of failed attempts.

\myparagraph{Improving Transferability}
We found that our attack weakly transfers to different networks in the black-box case, which could limit the effectiveness of the attack in practice.
Recently, a number of works have looked at the transferability of adversarial examples, providing insights on how to improve the generalization of such examples to unseen models.
One way to increase transferability is re-defining the optimization so that the adversarial perturbations can be applied across the entire distribution of legitimate input~\cite{Zheng2019, Xie2019, Zhao2019, Eykholt2017}.
A straightforward method to obtain this is to use the \textit{Expectation over Transformation} method (EOT) introduced in~\cite{Athalye2018}.
EOT has been used successfully in order to generate adversarial examples that are robust to real-world environmental changes.
It has also been shown that transferability is affected by the local smoothness of the loss function used for computing the perturbations~\cite{Wu2018}.
Enforcing such smoothness on the surrogate model during training leads to adversarial perturbations that better generalize to unseen models.

\section{Conclusion}
\label{sec:conclusion}




In this paper, we have presented a template poisoning attack that allows the adversary to place a biometric backdoor, which grants inconspicuous long-term access to the biometric system.
We designed the attack to cope with \knowledge and limited injection capabilities, showing that successful attacks can be obtained even in black-box models.
We investigated a set of recognition pipelines, including different models and classifiers.
We showed that some classifiers are particularly vulnerable to this attack, where a single poisoning sample injected can lead to success rates of over 40\%.
We suggested a new countermeasure that can successfully detect sets of consecutive poisoning samples based on their angular similarity in feature space.
We evaluated the trade-offs between poisoning detection and legitimate template updates being rejected, obtaining results of around 7\% EER for the detection on single sample, leading to $>$99\% attack detection rate after only two injected samples.
The weak assumptions of our attacker scenario and the increasing adoption of unsupervised template updates in deployed products highlight the severity of this attack.
These results suggest that increased attention should be given to the update procedure, as this represents an opportunity for attackers to compromise the authentication system.

\section*{Acknowledgements}\noindent
This work was generously supported by a grant from Mastercard and by the Engineering and Physical Sciences Research Council grant number EP/N509711/1.

\bibliographystyle{IEEEtran}
\balance
\bibliography{IEEEabrv,references}

\end{document}